\documentclass{emulateapj}
\usepackage{enumerate,multirow}

\shorttitle{Nuclear Star Cluster Scaling Relations}
\shortauthors{Scott \& Graham}

\begin{document}

\title{Updated Mass Scaling Relations for Nuclear Star Clusters and a Comparison to Supermassive Black Holes}
\author{Nicholas Scott and Alister W Graham}
\affil{Centre for Astrophysics and Supercomputing, Swinburne University of Technology, Hawthorn, Vic, 3122, Australia}

\begin{abstract}
We investigate whether nuclear star clusters and supermassive black holes follow a common set of mass scaling relations with their host galaxy's properties, and hence can be considered to form a single class of central massive object. We have compiled a large sample of galaxies with measured nuclear star cluster masses and host galaxy properties from the literature and fit log-linear scaling relations.  We find that nuclear star cluster mass, M$_{\rm NC}$, correlates most tightly with the host galaxy's velocity dispersion: log M$_{\rm NC} = (2.11\pm0.31) \log (\sigma/54) + (6.63\pm0.09)$, but has a slope dramatically shallower than the relation defined by supermassive black holes. We find that the nuclear star cluster mass relations involving host galaxy (and spheroid) luminosity and stellar and dynamical mass, intercept with but are in general shallower than the corresponding black hole scaling relations.  In particular M$_{\rm NC} \propto {\rm M}_{\rm Gal,dyn}^{0.55 \pm 0.15}$; the nuclear cluster mass is {\it not} a constant fraction of its host galaxy or spheroid mass. We conclude that nuclear stellar clusters and supermassive black holes do not form a single family of central massive objects.
\end{abstract}


\section{Introduction}

Central massive objects (CMOs) are a common feature in galaxies across the Hubble sequence. CMOs take the form of either a supermassive black hole (SMBH) or a compact stellar structure such as a nuclear stellar cluster (NC) or nuclear stellar disk (ND). The masses of SMBHs have been shown to correlate with a range of host galaxy properties including: stellar velocity dispersion, $\sigma$ \citep{Ferrarese:2000,Gebhardt:2000,Graham:2011}, stellar concentration \citep{Graham:2001,Graham:2007b}; dynamical mass, $M_\mathrm{dyn} \propto \sigma^2 R$ \citep{Magorrian:1998,Marconi:2003,Haring:2004,Graham:2012a}; and luminosity, L$_\mathrm{Sph}$ \citep{Kormendy:1995,Marconi:2003}.

Following the discovery that the luminosity of stellar CMOs correlates with that of their host bulge in disk galaxies \citep[][hereafter BGP07]{Balcells:2003,Balcells:2007} and elliptical galaxies \citep{Graham:2003}, the {\it masses} of stellar CMOs have also been shown to correlate with their host galaxy properties. NC mass has been reported to correlate with, for early-type galaxies, the host {\it galaxy's} luminosity, L$_\mathrm{Gal}$ and dynamical mass, as given by M$_\mathrm{Gal,dyn} \propto \sigma^2 R_{\rm e,Gal}$ \citep[][hereafter F06]{Ferrarese:2006}. Related correlations have also been reported with the host {\it spheroid's}: luminosity, L$_\mathrm{Sph}$ \citep[][hereafter WH06]{Wehner:2006}; stellar mass, M$_\mathrm{Sph,*}$ (BGP07); dynamical mass, M$_\mathrm{Sph,dyn}$ (WH06, BGP07) and velocity dispersion, $\sigma$ \citep[F06,][]{Graham:2012b}.

These scaling relations are physically interesting because they relate objects on very different scales: the gravitational sphere of influence of a SMBH is typically less than 0.1 per cent of its host galaxy's effective radius, R$_{\rm e}$. This connection is thought to be driven by feedback processes from the CMO \citep[e.g.][]{Silk:1998,Croton:2006,Booth:2009}, but may instead be related by the initial central stellar density of the host spheroid \citep{Graham:2007b}. While most studies have focused on feedback from black holes, analogous mechanisms driven by nuclear stellar clusters have been hypothesized \citep{McLaughlin:2006,McQuillin:2012}. One potential problem with these momentum-conserving feedback arguments, as constructed, is that they predict a slope of 4 for both the $M_{\rm BH}$--$\sigma$ and $M_{\rm NC}$--$\sigma$ relations, whereas the observations now suggest a slope of 5 \citep{Ferrarese:2000,Graham:2011} and somewhere between 1 and 2 \citep{Graham:2012b}, respectively. It should however be noted that the $\sigma$ term in the models relates to that of the dark matter halo rather than the stars, and as such they may not be appropriate for comparison.

F06 and WH06 have argued that SMBHs and NCs follow a single common scaling relation with M$_\mathrm{dyn}$ (though not with other host galaxy properties). Other investigations have reached different conclusions, for example  BGP07 find that NCs do not fall onto the linear relation defined by massive central black holes, and conclude that any CMO--bulge mass relation that encompasses both central black holes and nuclear star clusters must not be log-linear.

F06 have reported that the M$_\mathrm{NC}$--$\sigma$ relation has a slope which is consistent with the M$_\mathrm{BH}$--$\sigma$ relation.  However expanding upon the M$_\mathrm{CMO}$--$\sigma$ diagram from Figure 8 of \citet{Graham:2011}, \citet{Graham:2012b} has reported that $M_\mathrm{NC} \propto \sigma^1$ to $\sigma^2$, whereas $M_{\rm BH} \propto \sigma^5$ for non-barred galaxies\footnote{Barred galaxies tend to have higher velocity dispersions than given by the M$_\mathrm{BH}$--$\sigma$ relation defined by non-barred galaxies \citep{Graham:2008a,Hu:2008}. As such, the classical (i.e. all galaxy types) M$_\mathrm{BH} - \sigma$ relation has a slope $\sim 6$.} \citep{Ferrarese:2000,Graham:2011}. \citet{Leigh:2012} also report a significantly flatter slope of $M_\mathrm{NC} \propto \sigma^{2.73}$ for a sample of NCs.

The situation is complicated still further by a blurring of the division between galaxies containing SMBHs or NCs. Since F06, and WH06 who initially found a clear division in mass between galaxies hosting a SMBH (with M$_\mathrm{Sph,dyn} > 5\times10^9 M_\odot$) and galaxies hosting a NC (with M$_\mathrm{Sph,dyn} < 5\times10^9 M_\odot$), an increasing number of galaxies that host both a SMBH and a NC have been found \citep{Graham:2007b,Gonzalez:2008,Seth:2008}. \citet[][hereafter GS09]{Graham:2009} observed a transition region from $10^8 < {\rm M}_{\rm sph,*}/{\rm M}_{\odot} < 10^{10}$ where both types of nuclei coexist \citep[see also][]{Neumayer:2012}. These findings raise the question of how the combined CMO mass, M$_\mathrm{BH}$ + M$_\mathrm{NC}$, may scale with the host galaxy properties, though a larger sample of such objects is desired.

\citet{Graham:2012b} updated the $M_{\rm CMO}$--$\sigma$ diagram, first published by F06, using an expanded sample of galaxies with directly measured SMBH masses which also included 13 galaxies with both a NC and SMBH.  Here we re-examine and update the $M_{\rm CMO}$ versus (i) velocity dispersion, (ii) {\it B-}band galaxy magnitude and (iii) dynamical mass diagrams from F06.  In addition to the above sample expansion, we incorporate the NC data set from \citet{Balcells:2007}. We also construct another three diagrams involving M$_\mathrm{CMO}$ and: {\it K-}band luminosity; total stellar mass, M$_\mathrm{Gal,*}$; and spheroid stellar mass, M$_\mathrm{Sph,*}$.

Collectively our data represents the largest sample to date of NC and host galaxy properties. We have used this to investigate their scaling relations and whether they are consistent with those for SMBHs.  Our sample and data are more fully described in Section \ref{sec:sample}, while in Section \ref{sec:results} we present a range of NC and SMBH scaling relations. In Section \ref{sec:discussion} we discuss whether our results support the idea of a single common scaling relation for CMOs and the implications of our results on a common formation mechanism for SMBHs and NCs. Finally, we present a summary of our conclusions at the end of Section \ref{sec:discussion}.

\begin{table*}[t]
\caption{Nuclear star cluster, nuclear stellar disc and host galaxy properties}
\label{tab:sample}
\begin{center}
\begin{tabular*}{\textwidth}{@{\extracolsep{\fill}}l l c c c c c c c c c c c c c}
\hline
Galaxy & Type & m$_B$ & $ m-M $ & m$_{\rm NC}$ & M$_{\rm NC}$ & m$_{\rm ND}$ & M$_{\rm ND}$ & M/L & $\sigma$ & R$_e$ & M$_{\rm Gal,dyn}$ & m$_K$ & M$_{\rm Gal,*}$ & M$_{\rm Sph,*}$\\
 & & & & & log & & log & &  &  & log & & log & log \\
 & & mag & mag & mag & M$_\odot$ & mag & M$_\odot$ & & km\ s$^{-1}$ & (arcsec) & M$_\odot$ & mag & M$_\odot$ & M$_\odot$ \\
(1) & (2) & (3) & (4) & (5) & (6) & (7) & (8) & (9) & (10) & (11) & (12) & (13) & (14) & (15)\\
\hline
N4578 & S0 & 12.21$^\mathrm{[a]}$ & 31.06$^\mathrm{[d]}$ & 18.32$^\mathrm{[i]}$ & 7.59 & \dots & \dots & 2.80 & 124$^\mathrm{[j]}$ &  47$^\mathrm{[m]}$ & 10.8 & 8.40 & 10.3 & \dots \\
N4550 & E/S0 & 12.34$^\mathrm{[a]}$ & 30.95$^\mathrm{[d]}$ & \dots & \dots & 16.98$^\mathrm{[i]}$ & 8.09 & 2.87 & 116$^\mathrm{[j]}$ &  15$^\mathrm{[m]}$ & 10.2 & 8.69 & 10.1 & 10.1 \\
N4612 & S0 & 12.48$^\mathrm{[a]}$ & 31.10$^\mathrm{[d]}$ & 18.65$^\mathrm{[i]}$ & 7.18 & \dots & \dots & 1.42 &  84$^\mathrm{[j]}$ &  36$^\mathrm{[m]}$ & 10.4 & 8.56 & 10.3 & \dots \\
N4474 & S0 & 12.43$^\mathrm{[a]}$ & 30.96$^\mathrm{[d]}$ & 19.68$^\mathrm{[i]}$ & 7.09 & \dots & \dots & 3.40 &  44$^\mathrm{[j]}$ &  24$^\mathrm{[m]}$ & 9.7 & 8.70 & 10.1 & \dots \\
N4379 & S0 & 12.59$^\mathrm{[a]}$ & 31.00$^\mathrm{[d]}$ & 18.25$^\mathrm{[i]}$ & 7.66 & \dots & \dots & 3.28 & 115$^\mathrm{[j]}$ &  21$^\mathrm{[m]}$ & 10.4 & 8.77 & 10.1 & \dots \\
N4387 & E & 12.72$^\mathrm{[a]}$ & 31.27$^\mathrm{[d]}$ & 18.40$^\mathrm{[i]}$ & 7.54 & \dots & \dots & 2.23 & 108$^\mathrm{[j]}$ &  16$^\mathrm{[m]}$ & 10.2 & 9.15 & 10.1 & 10.1 \\
N4476 & S0 & 12.81$^\mathrm{[a]}$ & 31.23$^\mathrm{[d]}$ & 19.62$^\mathrm{[i]}$ & 7.05 & \dots & \dots & 2.29 &  29$^\mathrm{[j]}$ &  23$^\mathrm{[m]}$ & 10.0 & 9.46 & 9.9 & \dots \\
N4551 & E & 12.76$^\mathrm{[a]}$ & 31.04$^\mathrm{[d]}$ & \dots & \dots & 17.24$^\mathrm{[i]}$ & 8.06 & 3.11 & 101$^\mathrm{[j]}$ &  19$^\mathrm{[m]}$ & 10.3 & 8.87 & 10.1 & 10.1 \\
N4458 & E & 12.86$^\mathrm{[a]}$ & 31.07$^\mathrm{[d]}$ & \dots & \dots & 15.28$^\mathrm{[i]}$ & 8.72 & 2.30 & 101$^\mathrm{[j]}$ &  30$^\mathrm{[m]}$ & 10.4 & 9.31 & 9.9 & 9.9 \\
N4623 & E & 13.16$^\mathrm{[a]}$ & 31.20$^\mathrm{[d]}$ & \dots & \dots & 17.47$^\mathrm{[i]}$ & 7.95 & 2.55 &  71$^\mathrm{[j]}$ &  21$^\mathrm{[m]}$ & 10.0 & 9.47 & 9.9 & 9.9 \\
N4452 & S0 & 13.20$^\mathrm{[a]}$ & 31.09$^\mathrm{[e]}$ & 20.37$^\mathrm{[i]}$ & 6.38 & \dots & \dots & 1.10 &  67$^\mathrm{[j]}$ &  19$^\mathrm{[m]}$ & 9.8 & 9.10 & 10.0 & \dots \\
N4479 & S0 & 13.36$^\mathrm{[a]}$ & 31.20$^\mathrm{[d]}$ & 20.54$^\mathrm{[i]}$ & 6.70 & \dots & \dots & 2.44 &  64$^\mathrm{[j]}$ &  28$^\mathrm{[m]}$ & 10.0 & 9.77 & 9.8 & \dots \\
N4482 & dE & 13.47$^\mathrm{[a]}$ & 31.29$^\mathrm{[d]}$ & 19.39$^\mathrm{[i]}$ & 6.95 & \dots & \dots & 1.39 &   \dots &  30$^\mathrm{[m]}$ & \dots & 10.6 & 9.5 & 9.5 \\
N4352 & S0 & 13.53$^\mathrm{[a]}$ & 31.36$^\mathrm{[d]}$ & 19.83$^\mathrm{[i]}$ & 6.84 & \dots & \dots & 1.52 &  95$^\mathrm{[j]}$ &  23$^\mathrm{[m]}$ & 10.3 & 9.87 & 9.8 & \dots \\
I3468 & E & 13.55$^\mathrm{[a]}$ & 30.93$^\mathrm{[d]}$ & 20.10$^\mathrm{[i]}$ & 6.66 & \dots & \dots & 1.91 &  38$^\mathrm{[j]}$ &  28$^\mathrm{[m]}$ & 9.5 & 10.51 & 9.4 & 9.4 \\
I3773 & dS0 & 13.72$^\mathrm{[a]}$ & 31.09$^\mathrm{[e]}$ & 21.33$^\mathrm{[i]}$ & 6.10 & \dots & \dots & 1.41 &   \dots &  17$^\mathrm{[m]}$ & \dots & 10.9 & 9.3 & \dots \\
I3653 & E & 13.78$^\mathrm{[a]}$ & 30.95$^\mathrm{[d]}$ & 18.62$^\mathrm{[i]}$ & 7.28 & \dots & \dots & 1.99 &  54$^\mathrm{[j]}$ &  10$^\mathrm{[m]}$ & 9.4 & 10.58 & 9.4 & 9.4 \\
I809 & dE & 14.09$^\mathrm{[a]}$ & 31.03$^\mathrm{[d]}$ & 19.70$^\mathrm{[i]}$ & 6.79 & \dots & \dots & 1.64 &  43$^\mathrm{[j]}$ &  18$^\mathrm{[m]}$ & 9.7 & 10.61 & 9.4 & 9.4 \\
I3328 & dE & 14.20$^\mathrm{[a]}$ & 31.13$^\mathrm{[d]}$ & 18.88$^\mathrm{[i]}$ & 7.07 & \dots & \dots & 1.33 &   \dots &  21$^\mathrm{[m]}$ & \dots & 11.3 & 9.2 & 9.2 \\
I3065 & S0 & 14.20$^\mathrm{[a]}$ & 31.07$^\mathrm{[d]}$ & 21.95$^\mathrm{[i]}$ & 5.70 & \dots & \dots & 1.01 &  40$^\mathrm{[j]}$ &  13$^\mathrm{[m]}$ & 9.3 & 10.95 & 9.3 & \dots \\
I3442 & dE & 14.22$^\mathrm{[a]}$ & 31.14$^\mathrm{[d]}$ & 20.97$^\mathrm{[i]}$ & 6.17 & \dots & \dots & 1.12 &  25$^\mathrm{[j]}$ &  35$^\mathrm{[m]}$ & 9.3 & \dots & \dots & \dots \\
I3381 & dE & 14.25$^\mathrm{[a]}$ & 31.11$^\mathrm{[d]}$ & 20.12$^\mathrm{[i]}$ & 6.73 & \dots & \dots & 1.93 &  39$^\mathrm{[j]}$ &  27$^\mathrm{[m]}$ & 9.4 & 11.05 & 9.3 & 9.3 \\
I3652 & dE & 14.30$^\mathrm{[a]}$ & 31.04$^\mathrm{[d]}$ & 20.00$^\mathrm{[i]}$ & 6.57 & \dots & \dots & 1.27 &  27$^\mathrm{[j]}$ &  22$^\mathrm{[m]}$ & 9.2 & 11.05 & 9.2 & 9.2 \\
U7436 & dE & 14.31$^\mathrm{[a]}$ & 30.98$^\mathrm{[d]}$ & 22.44$^\mathrm{[i]}$ & 5.81 & \dots & \dots & 2.24 &   \dots &  24$^\mathrm{[m]}$ & \dots & 11.3 & 9.1 & 9.1 \\
I3470 & dE & 14.35$^\mathrm{[a]}$ & 31.04$^\mathrm{[d]}$ & 19.47$^\mathrm{[i]}$ & 6.80 & \dots & \dots & 1.33 &  50$^\mathrm{[j]}$ &  14$^\mathrm{[m]}$ & 9.5 & 11.18 & 9.2 & 9.2 \\
I3501 & dE & 14.45$^\mathrm{[a]}$ & 31.06$^\mathrm{[d]}$ & 22.16$^\mathrm{[i]}$ & 5.66 & \dots & \dots & 1.12 &  35$^\mathrm{[j]}$ &  14$^\mathrm{[m]}$ & 9.2 & 11.24 & 9.2 & 9.2 \\
I3586 & dS0 & 14.40$^\mathrm{[a]}$ & 31.09$^\mathrm{[e]}$ & 22.44$^\mathrm{[i]}$ & 5.81 & \dots & \dots & 1.99 &  26$^\mathrm{[j]}$ &  29$^\mathrm{[m]}$ & 9.3 & 12.13 & 8.8 & \dots \\
U7399A & dE & 14.48$^\mathrm{[a]}$ & 31.17$^\mathrm{[d]}$ & 19.89$^\mathrm{[i]}$ & 6.59 & \dots & \dots & 1.07 &  41$^\mathrm{[j]}$ &  36$^\mathrm{[m]}$ & 9.8 & 11.53 & 9.1 & 9.1 \\
I3735 & dE & 14.52$^\mathrm{[a]}$ & 31.16$^\mathrm{[d]}$ & 20.23$^\mathrm{[i]}$ & 6.54 & \dots & \dots & 1.31 &  41$^\mathrm{[j]}$ &  22$^\mathrm{[m]}$ & 9.6 & 11.38 & 9.1 & 9.1 \\
I3032 & dE & 14.57$^\mathrm{[a]}$ & 30.89$^\mathrm{[d]}$ & 22.04$^\mathrm{[i]}$ & 5.66 & \dots & \dots & 1.19 &   \dots &  13$^\mathrm{[m]}$ & \dots & 12.17 & 8.7 & 8.7 \\
V200 & dE & 14.63$^\mathrm{[a]}$ & 31.30$^\mathrm{[d]}$ & 22.75$^\mathrm{[i]}$ & 5.40 & \dots & \dots & 0.85 &   \dots &  18$^\mathrm{[m]}$ & \dots & 12.51 & 8.7 & 8.7 \\
I3487 & E & 14.74$^\mathrm{[a]}$ & 31.09$^\mathrm{[e]}$ & 23.63$^\mathrm{[i]}$ & 5.00 & \dots & \dots & 0.92 &   \dots &  13$^\mathrm{[m]}$ & \dots & 12.22 & 8.8 & 8.8 \\
U7854 & dE & 14.91$^\mathrm{[a]}$ & 31.00$^\mathrm{[d]}$ & 23.42$^\mathrm{[i]}$ & 5.12 & \dots & \dots & 1.10 &   \dots &  14$^\mathrm{[m]}$ & \dots & 12.31 & 8.7 & 8.7 \\
I3509 & E & 14.85$^\mathrm{[a]}$ & 31.13$^\mathrm{[d]}$ & 21.77$^\mathrm{[i]}$ & 5.85 & \dots & \dots & 1.14 &   \dots &  16$^\mathrm{[m]}$ & \dots & 11.98 & 8.9 & 8.9 \\
N4467 & E & 15.01$^\mathrm{[a]}$ & 31.09$^\mathrm{[e]}$ & 19.00$^\mathrm{[i]}$ & 7.09 & \dots & \dots & 1.61 &  68$^\mathrm{[j]}$ &  11$^\mathrm{[m]}$ & 9.2 & 10.49 & 9.5 & 9.5 \\
I3383 & dE & 15.04$^\mathrm{[a]}$ & 31.04$^\mathrm{[d]}$ & 20.97$^\mathrm{[i]}$ & 6.14 & \dots & \dots & 1.15 &  37$^\mathrm{[j]}$ &  21$^\mathrm{[m]}$ & 9.3 & 12.82 & 8.5 & 8.5 \\
V1627 & E & 15.07$^\mathrm{[a]}$ & 30.97$^\mathrm{[d]}$ & 18.68$^\mathrm{[i]}$ & 7.35 & \dots & \dots & 2.46 &   \dots &   6$^\mathrm{[m]}$ & \dots & 11.66 & 9.0 & 9.0 \\
I798 & E & 15.16$^\mathrm{[a]}$ & 31.02$^\mathrm{[d]}$ & 19.59$^\mathrm{[i]}$ & 6.93 & \dots & \dots & 2.07 &   \dots &  11$^\mathrm{[m]}$ & \dots & 11.3 & 9.0 & 9.0 \\
I3101 & dE & 15.16$^\mathrm{[a]}$ & 31.25$^\mathrm{[d]}$ & 20.20$^\mathrm{[i]}$ & 6.52 & \dots & \dots & 1.12 &   \dots &  14$^\mathrm{[m]}$ & \dots & 12.92 & 8.6 & 8.6 \\
I3779 & dE & 15.18$^\mathrm{[a]}$ & 30.99$^\mathrm{[d]}$ & 22.29$^\mathrm{[i]}$ & 5.62 & \dots & \dots & 1.23 &   \dots &  15$^\mathrm{[m]}$ & \dots & 12.81 & 8.5 & 8.5 \\
I3292 & dS0 & 15.24$^\mathrm{[a]}$ & 30.99$^\mathrm{[d]}$ & 21.10$^\mathrm{[i]}$ & 6.13 & \dots & \dots & 1.33 &   \dots &  14$^\mathrm{[m]}$ & \dots & 11.83 & 8.9 & \dots \\
I3635 & dE & 15.25$^\mathrm{[a]}$ & 31.13$^\mathrm{[d]}$ & 21.36$^\mathrm{[i]}$ & 5.99 & \dots & \dots & 1.08 &   \dots &  21$^\mathrm{[m]}$ & \dots & 12.74 & 8.6 & 8.6 \\
N4309A & E & 15.40$^\mathrm{[a]}$ & 31.80$^\mathrm{[d]}$ & 21.19$^\mathrm{[i]}$ & 6.16 & \dots & \dots & 0.75 &   \dots &   7$^\mathrm{[m]}$ & \dots & 13.28 & 8.6 & 8.6 \\
I3461 & dE & 15.44$^\mathrm{[a]}$ & 31.12$^\mathrm{[d]}$ & 20.27$^\mathrm{[i]}$ & 6.42 & \dots & \dots & 1.08 &   \dots &  15$^\mathrm{[m]}$ & \dots & 12.41 & 8.7 & 8.7 \\
V1886 & dE & 15.43$^\mathrm{[a]}$ & 31.09$^\mathrm{[e]}$ & 21.92$^\mathrm{[i]}$ & 5.77 & \dots & \dots & 1.15 &  37$^\mathrm{[j]}$ &  18$^\mathrm{[m]}$ & 9.5 & \dots & \dots & \dots \\
V1199 & E & 15.49$^\mathrm{[a]}$ & 31.09$^\mathrm{[e]}$ & 19.67$^\mathrm{[i]}$ & 6.94 & \dots & \dots & 2.09 &  61$^\mathrm{[j]}$ &   5$^\mathrm{[m]}$ & 9.9 & 12.34 & 8.7 & 8.7 \\
V1539 & dE & 15.63$^\mathrm{[a]}$ & 31.14$^\mathrm{[d]}$ & 20.81$^\mathrm{[i]}$ & 6.28 & \dots & \dots & 1.26 &   \dots &  38$^\mathrm{[m]}$ & \dots & \dots & \dots & \dots \\
V1185 & dE & 15.67$^\mathrm{[a]}$ & 31.14$^\mathrm{[d]}$ & 20.77$^\mathrm{[i]}$ & 6.21 & \dots & \dots & 1.04 &   \dots &  27$^\mathrm{[m]}$ & \dots & 13.26 & 8.4 & 8.4 \\
I3633 & dE & 15.72$^\mathrm{[a]}$ & 31.05$^\mathrm{[d]}$ & 20.04$^\mathrm{[i]}$ & 6.69 & \dots & \dots & 1.72 &   \dots &  10$^\mathrm{[m]}$ & \dots & 13.50 & 8.3 & 8.3 \\
I3490 & dE & 15.84$^\mathrm{[a]}$ & 31.09$^\mathrm{[e]}$ & 22.25$^\mathrm{[i]}$ & 5.56 & \dots & \dots & 0.96 &   \dots &  15$^\mathrm{[m]}$ & \dots & \dots & \dots & \dots \\
V1661 & dE & 15.98$^\mathrm{[a]}$ & 31.00$^\mathrm{[d]}$ & 20.22$^\mathrm{[i]}$ & 6.49 & \dots & \dots & 1.37 &   \dots &  84$^\mathrm{[m]}$ & \dots & \dots & \dots & \dots \\
M32 & cE & 8.76$^\mathrm{[b]}$ & 24.49$^\mathrm{[f]}$ & \dots & 7.30$^\mathrm{[f]}$ & \dots & \dots & \dots &  72$^\mathrm{[k]}$ & 12$^\mathrm{[b]}$ & 8.9 & 5.09 & 9.0 & 8.41 \\
N1023 & SB0 & 10.09$^\mathrm{[b]}$ & 30.23$^\mathrm{[f]}$ & \dots & 6.64$^\mathrm{[f]}$ & \dots & \dots & \dots & 204$^\mathrm{[k]}$ & 13$^\mathrm{[b]}$ & 11.0 & 6.24 & 10.8 & 10.5 \\
N1399 & E & 10.49$^\mathrm{[b]}$ & 31.44$^\mathrm{[f]}$ & \dots & 6.81$^\mathrm{[f]}$ & \dots & \dots & \dots & 329$^\mathrm{[k]}$ & 13$^\mathrm{[b]}$ & 11.7 & 6.31 & 11.3 & 11.2 \\
N2778 & SB0 & 13.26$^\mathrm{[b]}$ & 31.74$^\mathrm{[f]}$ & \dots & 6.83$^\mathrm{[f]}$ & \dots & \dots & \dots & 162$^\mathrm{[k]}$ & 5$^\mathrm{[b]}$ & 10.7 & 9.51 & 10.1 & 9.6 \\
N3115 & S0 & 9.67$^\mathrm{[b]}$ & 29.87$^\mathrm{[f]}$ & \dots & 7.18$^\mathrm{[f]}$ & \dots & \dots & \dots & 252$^\mathrm{[k]}$ & 11$^\mathrm{[b]}$ & 11.0 & 5.88 & 10.9 & 10.9 \\
N3384 & SB0 & 10.73$^\mathrm{[b]}$ & 30.27$^\mathrm{[f]}$ & \dots & 7.34$^\mathrm{[f]}$ & \dots & \dots & \dots & 148$^\mathrm{[k]}$ & 8$^\mathrm{[b]}$ & 10.5 & 6.75 & 10.6 & 10.2 \\
N4026 & S0 & 11.58$^\mathrm{[b]}$ & 30.60$^\mathrm{[f]}$ & \dots & 7.11$^\mathrm{[f]}$ & \dots & \dots & \dots & 178$^\mathrm{[k]}$ & 5$^\mathrm{[b]}$ & 10.6 & 7.58 & 10.4 & 10.0 \\
N4395 & Sm & 10.57$^\mathrm{[b]}$ & 28.07$^\mathrm{[f]}$ & \dots & 6.15$^\mathrm{[f]}$ & \dots & \dots & \dots &   \dots & 58$^\mathrm{[b]}$ & \dots & 10.0 & 8.5 & 7.5 \\
N4697 & E & 10.01$^\mathrm{[b]}$ & 30.33$^\mathrm{[f]}$ & \dots & 7.45$^\mathrm{[f]}$ & \dots & \dots & \dots & 171$^\mathrm{[k]}$ & 24$^\mathrm{[b]}$ & 11.1 & 6.37 & 11.1 & 11.1 \\
M33 & Scd & 6.09$^\mathrm{[b]}$ & 24.72$^\mathrm{[f]}$ & \dots & 6.30$^\mathrm{[f]}$ & \dots & \dots & \dots &  37$^\mathrm{[k]}$ & 269$^\mathrm{[b]}$ & \dots & 4.10 & 9.5 & 8.2 \\
N205 & E & 8.65$^\mathrm{[b]}$ & 24.52$^\mathrm{[f]}$ & \dots & 6.15$^\mathrm{[f]}$ & \dots & \dots & \dots &  20$^\mathrm{[k]}$ & 56$^\mathrm{[b]}$ & 8.5 & 5.59 & 8.9 & 8.9 \\
\hline
\end{tabular*}
\end{center}
\end{table*}
\begin{table*}[t]
\begin{center}
{\sc Table 1 continued}
\begin{tabular*}{\textwidth}{@{\extracolsep{\fill}}l l c c c c c c c c c c c c c}
\hline
Galaxy & Type & m$_B$ & $m$-$M$ & m$_{\rm NC}$ & M$_{\rm NC}$ & m$_{\rm ND}$ & M$_{\rm ND}$ & M/L & $\sigma$ & R$_e$ & M$_{\rm dyn}$ & m$_K$ & M$_{\rm Gal}$ & M$_{\rm Sph}$\\
 & & & & & log & & log & &  &  & log & & log & log \\
 & & mag & mag & mag & M$_\odot$ & mag & M$_\odot$ & & km\ s$^{-1}$ & arcsec & M$_\odot$ & mag & M$_\odot$ & M$_\odot$ \\
(1) & (2) & (3) & (4) & (5) & (6) & (7) & (8) & (9) & (10) & (11) & (12) & (13) & (14) & (15)\\
\hline
N3621 & Sd & 9.93$^\mathrm{[b]}$ & 29.16$^\mathrm{[f]}$ & \dots & 7.00$^\mathrm{[f]}$ & \dots & \dots & \dots &   \dots &  27$^\mathrm{[b]}$ & \dots & 6.60 & 10.2 & 8.2 \\
N4041 & Sbc & 11.80$^\mathrm{[b]}$ & 31.78$^\mathrm{[f]}$ & \dots & 7.46$^\mathrm{[f]}$ & \dots & \dots & \dots &   \dots &   \dots & \dots & 8.41 & 10.6 & 8.8 \\
V1254 & dE & 15.02$^\mathrm{[b]}$ & 31.09$^\mathrm{[e]}$ & \dots & 7.04$^\mathrm{[f]}$ & \dots & \dots & \dots &  31$^\mathrm{[k]}$ &   4$^\mathrm{[b]}$ & 9.0 & 11.70 & 9.5 & 9.5 \\
N1300 & SBbc & 10.98$^\mathrm{[b]}$ & 31.58$^\mathrm{[g]}$ & \dots & 7.94$^\mathrm{[g]}$ & \dots & \dots & \dots & 229$^\mathrm{[k]}$ &  33$^\mathrm{[b]}$ & \dots & 7.56 & 10.8 & \dots \\
N2549 & SB0 & 11.91$^\mathrm{[b]}$ & 30.45$^\mathrm{[g]}$ & \dots & 7.04$^\mathrm{[g]}$ & \dots & \dots & \dots & 144$^\mathrm{[k]}$ &   6$^\mathrm{[b]}$ & 10.4 & 8.05 & 10.2 & \dots \\
N3585 & S0 & 10.60$^\mathrm{[b]}$ & 31.45$^\mathrm{[g]}$ & \dots & 6.60$^\mathrm{[g]}$ & \dots & \dots & \dots & 206$^\mathrm{[k]}$ &  12$^\mathrm{[b]}$ & 11.2 & 6.70 & 11.1 & \dots \\
Milky & \multirow{2}{*}{\dots} & \multirow{2}{*}{-5.78$^\mathrm{[c]}$} & \multirow{2}{*}{14.52$^\mathrm{[c]}$} & \multirow{2}{*}{\dots} & \multirow{2}{*}{7.48$^\mathrm{[f]}$} & \multirow{2}{*}{\dots} 
& \multirow{2}{*}{\dots} & \multirow{2}{*}{\dots} & \multirow{2}{*}{100$^\mathrm{[k]}$} & \multirow{2}{*}{33884$^\mathrm{[c]}$} & \multirow{2}{*}{10.7} & \multirow{2}{*}{-9.27} & \multirow{2}{*}{10.7} & \multirow{2}{*}{10.1} \\
Way & & & & & & & & & & & & & & \\
N5326 & Sa & 12.83$^\mathrm{[b]}$ & 32.68$^\mathrm{[h]}$ & \dots & \dots & 13.47$^\mathrm{[h]}$ & 8.90 & 0.80 & 164$^\mathrm{[l]}$ &  12$^\mathrm{[b]}$ & 11.3 & 8.88 & 10.7 & 10.4 \\
N5389 & S0 & 12.79$^\mathrm{[b]}$ & 32.09$^\mathrm{[h]}$ & \dots & \dots & 15.02$^\mathrm{[h]}$ & 8.04 & 0.80 & 114$^\mathrm{[l]}$ &   6$^\mathrm{[b]}$ & 10.6 & 8.62 & 10.6 & 10.1 \\
N5422 & S0 & 12.75$^\mathrm{[b]}$ & 32.02$^\mathrm{[h]}$ & 17.23$^\mathrm{[h]}$ & 7.13 & \dots & \dots & 0.80 & 160$^\mathrm{[l]}$ &   \dots & \dots & 8.76 & 10.5 & 10.0 \\
N5443 & SBb & 13.29$^\mathrm{[b]}$ & 32.17$^\mathrm{[h]}$ & \dots & \dots & 16.18$^\mathrm{[h]}$ & 7.61 & 0.80 &  76$^\mathrm{[l]}$ &   \dots & \dots & 9.04 & 10.5 & 9.5 \\
N5475 & Sa & 13.37$^\mathrm{[b]}$ & 31.95$^\mathrm{[h]}$ & 16.95$^\mathrm{[h]}$ & 7.22 & 13.25$^\mathrm{[h]}$ & 8.70 & 0.80 &  91$^\mathrm{[l]}$ &   \dots & \dots & 9.40 & 10.2 & 9.6 \\
N5587 & S0 & 13.74$^\mathrm{[b]}$ & 32.46$^\mathrm{[h]}$ & \dots & \dots & 14.23$^\mathrm{[h]}$ & 8.51 & 0.80 &  93$^\mathrm{[l]}$ &   \dots & \dots & 9.68 & 10.3 & 9.3 \\
N5689 & S0 & 12.65$^\mathrm{[b]}$ & 32.41$^\mathrm{[h]}$ & \dots & \dots & 13.13$^\mathrm{[h]}$ & 8.93 & 0.80 & 143$^\mathrm{[l]}$ &   \dots & \dots & 8.40 & 10.8 & 10.1 \\
N5707 & Sab & 13.23$^\mathrm{[b]}$ & 32.46$^\mathrm{[h]}$ & 16.30$^\mathrm{[h]}$ & 7.68 & 12.69$^\mathrm{[h]}$ & 9.12 & 0.80 & 141$^\mathrm{[l]}$ &   \dots & \dots & \dots & \dots & 10.0 \\
N5719 & SBa & 13.14$^\mathrm{[b]}$ & 31.82$^\mathrm{[h]}$ & \dots & \dots & 15.18$^\mathrm{[h]}$ & 7.87 & 0.80 & 108$^\mathrm{[l]}$ &  10$^\mathrm{[b]}$ & 10.7 & 8.23 & 10.7 & 10.4 \\
N5746 & SBb & 11.12$^\mathrm{[b]}$ & 31.80$^\mathrm{[h]}$ & 17.08$^\mathrm{[h]}$ & 7.10 & \dots & \dots & 0.80 & 139$^\mathrm{[l]}$ &  25$^\mathrm{[b]}$ & \dots & 6.88 & 11.2 & 10.1 \\
N5838 & E/S0 & 11.69$^\mathrm{[b]}$ & 31.31$^\mathrm{[h]}$ & 15.18$^\mathrm{[h]}$ & 7.67 & 11.15$^\mathrm{[h]}$ & 9.28 & 0.80 & 255$^\mathrm{[l]}$ & 8$^\mathrm{[b]}$ & 11.2 & 7.58 & 10.7 & 10.3 \\
N5854 & S0 & 12.48$^\mathrm{[b]}$ & 31.84$^\mathrm{[h]}$ & 14.94$^\mathrm{[h]}$ & 7.98 & 12.70$^\mathrm{[h]}$ & 8.87 & 0.80 &  97$^\mathrm{[l]}$ & 5$^\mathrm{[b]}$ & 10.3 & 8.82 & 10.4 & 9.9 \\
N5879 & Sbc & 12.17$^\mathrm{[b]}$ & 30.70$^\mathrm{[h]}$ & 17.42$^\mathrm{[h]}$ & 6.53 & \dots & \dots & 0.80 &  58$^\mathrm{[l]}$ & 9$^\mathrm{[b]}$ & \dots & 8.79 & 10.0 & 9.0 \\
N6010 & S0 & 12.69$^\mathrm{[b]}$ & 32.07$^\mathrm{[h]}$ & 16.29$^\mathrm{[h]}$ & 7.53 & \dots & \dots & 0.80 & 144$^\mathrm{[l]}$ & \dots & \dots & 8.93 & 10.5 & 9.9 \\
N6504 & Sa & 13.06$^\mathrm{[b]}$ & 33.96$^\mathrm{[h]}$ & 17.25$^\mathrm{[h]}$ & 7.90 & \dots & \dots & 0.80 & 185$^\mathrm{[l]}$ & \dots & \dots & 9.22 & 11.1 & 10.7 \\
N7457 & S0 & 11.87$^\mathrm{[b]}$ & 30.68$^\mathrm{[h]}$ & 14.97$^\mathrm{[h]}$ & 7.50 & 14.39$^\mathrm{[h]}$ & 7.73 & 0.80 &  56$^\mathrm{[l]}$ & 11$^\mathrm{[b]}$ & 9.89 & 8.19 & 10.2 & 9.2 \\
N7537 & Sbc & 13.57$^\mathrm{[b]}$ & 32.74$^\mathrm{[h]}$ & 17.40$^\mathrm{[h]}$ & 7.35 & \dots & \dots & 0.80 &  42$^\mathrm{[l]}$ & 5$^\mathrm{[b]}$ & \dots & 10.21 & 10.2 & 9.0 \\
\hline
Totals: & 86 & 86 & 86 & 57 & 76 & 15 & 15 & 68 & 61 & 77 & 48 & 80 & 80 & 66 \\
\hline
\end{tabular*}
\end{center}
\begin{minipage}{17.8cm}
Notes: Column (1): Galaxy ID. Column (2): Morphological type. Column (3): Galaxy {\it B-}band magnitude. Column (4): Distance modulus. Column (5): Nuclear star cluster magnitude. Column (6): Nuclear star cluster stellar mass. Column (7): Nuclear stellar disk magnitude. Column (8): Nuclear stellar disk stellar mass. Column (9) CMO mass-to-light ratio, in the band appropriate to the magnitudes given in Column (5) and Column (7). Column (10) Galaxy velocity dispersion. Column (11): Galaxy effective radius. Column (12): Galaxy dynamical mass. Column (13): Galaxy {\it K-}band magnitude. Column (14): Galaxy stellar mass. Column (15): Spheroid stellar mass.\\
References: [a] \citet{Binggeli:1985}. [b] RC3. [c] \citet{Cardone:2005} [d] \citet{Mei:2007}. [e] Assigned the median distance for Virgo from \citet{Mei:2007}. [f] GS09. [g] \citet{Graham:2012b}. [h] BGP07. All magnitudes are $K-$band. [i] $g-$band magnitudes from \citet{Cote:2006}. [j] F06. [k] \citet{Graham:2011}. [l] \citet{Falcon:2002}. [m] \citet{Ferrarese:2006}.
\end{minipage}
\end{table*}

\section{Sample and Data}
\label{sec:sample}
We constructed our sample of nuclear stellar objects by combining the data from F06 (51 objects), BGP07 (17 objects) and GS09 (16 objects). \citet{Graham:2012b} added a further 3 objects to the GS09 sample for a total of 19 objects, 15 with both a NC and a SMBH and a further 4 objects with a NC and only an upper limit on M$_\mathrm{BH}$. NGC7457 appears in both the BGP07 and GS09 samples, however the GS09 nuclear cluster properties are taken directly from BGP07. We eliminate this duplicate galaxy, reducing our sample by one galaxy. This gives a final sample of 86 objects with measured $M_{\rm CMO}$. The observed and derived properties of the nuclear star clusters and their host galaxies are described in full in the following sections, and are presented in Table \ref{tab:sample}.

\subsection{Nuclear stellar masses} 

GS09 provide stellar masses for their nuclei, whereas F06 and BGP07 tabulate only magnitudes. For the F06 objects we derived nuclear stellar object masses following F06. We multiplied the total CMO {\it g-}band magnitude by a mass-to-light ratio, $M/L$, determined from the single stellar population models of \citet{Bruzual:2003}, using the nuclear cluster colors given in \citet{Cote:2006} and a stellar population age $t=5$ Gyrs. For the BGP07 objects we derived masses following BGP07, by multiplying the total CMO {\it K}-band magnitude by $M/L_K=0.8$ \citep{Bell:2001} based on typical colors of the bulge population. The uncertainties on the nuclear object masses for the F06, BGP07 and GS09 data are given by the respective authors as 45 per cent, 33 per cent and a factor of 2 respectively. In passing we note that if NCs are related to ultra compact dwarf galaxies \citep[e.g.][]{Kroupa:2010} they may have a high stellar $M/L$ due to either a bottom-heavy \citep{Mieske:2008} or top-heavy initial mass function \citep{Dabringhausen:2009}.

BGP07 distinguish between extended nuclear components (11 objects) and unresolved nuclear components (also 11 objects -- 5 galaxies contain both a resolved and unresolved nuclear component), finding that the extended components are well fit with an exponential profile and are thus likely to be nuclear disks (or possibly nuclear bars), whereas the unresolved components are probably nuclear star clusters. They revealed that the disks and clusters follow quite different relations, in terms of, for example, how the nuclear disk luminosity scales with host galaxy $\sigma$. For this reason it is important to distinguish between nuclear disks and clusters when examining the scaling relations of nuclear objects with their host galaxy.

F06 identified three of their objects as containing small-scale stellar disks. Based on their published {\it HST} surface photometry we identify one further object, NGC 4550 (VCC 1619) as likely containing a nuclear stellar disk. These four nuclear disks are also the most extended nuclear components in the F06 sample, having half-light radii ranging from 26 to 63 pc (for comparison, the mean half-light radii for the F06 nuclear objects is 4 pc). This provides final samples of 76 and 15 nuclear clusters and nuclear disks, respectively, from the galaxy samples of F06, BGP07 and GS09. 5 galaxies contain both a nuclear star cluster and a nuclear disk, thus 86 unique galaxies with a stellar CMO.

\subsection{Host galaxy and spheroid properties}

We we were not able to obtain every galaxy and spheroid property for every object from the literature. The number of objects for which we were able to obtain a given property is indicated in the final row of Table \ref{tab:sample}. Velocity dispersions were obtained from F06, \citet[for the BGP07 galaxies]{Falcon:2002} and \citet[for the GS09 galaxies]{Graham:2011}, giving 51/76 nuclear star cluster galaxies and 15/15 nuclear disk galaxies having measured $\sigma$. The velocity dispersions were measured in inhomogeneous apertures and we do not attempt to correct the measurements to a common aperture here. The F06 $\sigma$ values were measured from long-slit observations within a 1 R$_e$ aperture. The values from \citet{Falcon:2002} were obtained within a central 1.1 arcsec$^2$ aperture and corrected to a `standard' aperture defined to be equivalent to a circular aperture with radius 1.7 arcsec at the distance of Coma \citep[as established by][]{Jorgensen:1995}. The $\sigma$ values presented in \citet{Graham:2011} were originally drawn from the HyperLeda database \citep{Paturel:2003}\footnote{http://leda.univ-lyon1.fr} and represent a disparate set of measurements corrected to the same `standard' aperture as in \citet{Falcon:2002}. We adopt an uncertainty of 10 \% for all $\sigma$ measurements.

We considered correcting the standard aperture measurements to R$_e$ measurements based on equation 1 of \citet{Cappellari:2006}, from which we derived a mean correction to the $\sigma$ measurements reported here of 2.6 \% (ranging from $+4$ \% to $-6$ \% for individual objects). However the error on the derived correction for individual objects is significant, $\sim 10 \%$, and much larger than the typical correction of $\sim 2.5 \%$ for a given measurement. The correction is not correlated with host galaxy $\sigma$, hence will not introduce a systematic error in our uncorrected $\sigma$ values. Given this uncertainty, and that we were unable to derive an aperture correction for all our objects due to missing R$_e$ measurements, we opted not to apply any aperture correction. 

We obtained total apparent {\it B}-band magnitudes for all galaxies following F06. For the BGP07 and GS09 galaxies we obtained $m_B$ from \citet[][RC3]{deVaucouleurs:1991}. For the F06 galaxies we obtained $m_B$ from \citet{Binggeli:1985}, reduced to the RC3 system using the relation given in the HyperLeda database. We note that this approach fails to fully correct for dust in disc galaxies \citep[see][]{Graham:2008b}. We therefore also obtained total {\it K-}band magnitudes, m$_K$, for 80/86 galaxies from the 2 Micron All Sky Survey (2MASS) Extended Source Catalogue \citep{Jarrett:2000}. To convert to absolute magnitudes we used the distances from \citet{Mei:2007} for the F06 sample, and from BGP07 and GS09 for the corresponding galaxies. We adopt an uncertainty of 0.25 mag for all absolute magnitudes.

We derive dynamical masses using the simple but popular virial estimator: M$_\mathrm{dyn} = \alpha \sigma_e^2 \mathrm{R}_e / \mathrm{G}$, where $R_{\rm e}$ is the effective half-light radius and $\sigma_e$ the luminosity-weighted velocity dispersion measured within a 1 R$_e$ aperture. Following F06, we used a value of $\alpha = 5$ for the F06 galaxy sample. The virial factor $\alpha$ can take on a range of values \citep{Bertin:2002} depending on the radial mass distribution. By comparing virial estimator derived masses to the results of more sophisticated dynamical models, \citet{Cappellari:2006} found that, in practical situations when working with real data, $\alpha=5$ provides a virtually unbiased estimate of a galaxy's dynamical mass within 1 R$_e$.. They found this to be true for galaxies with a broad range of S\'ersic indices, $n=2-10$ and bulge-to-total ratios, $B/T \sim 0.2-1.0$. They caution, however, that their result ``strictly applies to virial measurements derived... using `classic' determination of R$_e$ and L via R$^{1/4}$ growth curves, and with $\sigma_e$ measured in a large aperture." Based on their findings we conclude that the virial estimator is a reasonable approximation of the dynamical mass for galaxies of types Sa and earlier -- we do not determine M$_\mathrm{Gal,dyn}$ for galaxies of morphological type Sb or later, as these are heavily disk-dominated systems for which the virial estimator has not been calibrated.

\begin{table*}[t]
\caption{Nuclear cluster and black hole scaling relations}
\label{tab:fits}
\centering
\begin{tabular}{l l c c c c c c c r}
\hline
& Relation & a & err(a) & b & err(b) & $\sigma_{rms}$ & r  & N &Notes\\
&  (1)     & (2) &  (3)     & (4) &   (5)    &    (6)           & (7) & (8) &  (9)\\
\hline
\hline
\multirow{6}{*}{Figure \ref{fig:scaling1}}&M$_{B}$ + 16.9,log M$_{NC}$&6.44&0.06&-0.32&0.05&0.55&-0.64&76 &\\
&M$_{B}$ + 19.9,log M$_{BH}$&8.20&0.09&-0.65&0.11&0.39&-0.41&25 &\\
\cline{2-10}
&log $\sigma/54.0$,log M$_{NC}$&6.63&0.09&2.11&0.31&0.55&0.62&51 &\\
&log $\sigma/224.0$,log M$_{BH}$&8.46&0.06&6.10&0.44&0.47&0.88&64 &\\
\cline{2-10}
&log M$_{Gal,dyn}/10^{9.6}$,log M$_{NC}$&6.65&0.10&0.55&0.15&0.50&0.53&41 & Ex. Sb and later\\
&log M$_{Gal,dyn}/10^{11.3}$,log M$_{BH}$&8.47&0.07&1.37&0.23&0.46&0.76&40 &Ex. Sb and later\\
\hline
\multirow{6}{*}{Figure \ref{fig:scaling2}} &M$_{K}$ + 20.4,log M$_{NC}$&6.63&0.07&-0.24&0.04&0.52&-0.69&57 & E and dE only\\
&M$_{K}$ + 23.4,log M$_{BH}$&8.04&0.14&-0.48&0.09&0.40&-0.70&25 & E and dE only\\ 
\cline{2-10}
&log M$_{Gal,*}/10^{9.6}$,log M$_{NC}$&6.73&0.06&0.80&0.10&0.53&0.72&71 &\\
&log M$_{Gal,*}/10^{11.3}$,log M$_{BH}$&9.40&0.32&2.72&0.69&1.03&0.55&59 &\\
\cline{2-10}
&log M$_{Sph,*}/10^{9.6}$,log M$_{NC}$&7.02&0.10&0.88&0.19&0.63&0.64&57 &\\
&log M$_{Sph,*}/10^{11.3}$,log M$_{BH}$&8.80&0.11&1.20&0.19&0.63&0.65&39 &\\
\hline
 \end{tabular}
 \\
Column (1): $X$ and $Y$ parameters of the linear regression. 
Columns (2)-(5): Slope $b$ and zeropoint $a$, and their associated error, from the best-fitting linear relation. 
Column (6): Root mean square (rms) scatter in the $\log {\rm M}_{\rm CMO}$ direction.
Column(7): Spearman $r$ coefficient. 
Column (8): Number of data points contributing to the fit.\\
Fits of the form $\log y = a + b \log x$ (or $\log y = a + bx$
for $M_{B}$ and M$_K$), were performed using the {\sc BCES(Orth)} regression. 
\end{table*}

For the BGP07 and GS09 objects we use $R_{\rm e}$ values from the RC3 (which are determined from $R^{1/4}$ curve-of-growth fits to the surface brightness profile). For the F06 objects we use $R_{\rm e}$ values from \citet{Ferrarese:2006b} which are derived from S\'ersic R$^{1/n}$ fits to the observed surface brightness profile. For these 51 objects F06 report a range in $n$ from 0.8 to 4.6 (with 78 per cent of galaxies with $n$ in the range 1 to 2.5). We compared S\'ersic-based R$_e$,s for the subset of F06 galaxies for which RC3 R$^{1/4}$-based R$_{\rm e,deV}$ were available (22 objects) and found a one-to-one correlation, with the F06 R$_e$,s being systematically $30 \pm 7 \%$ larger than the R$_{\rm e,deV}$ values. For these comparison galaxies $n$ ranged from 1.1 to 4.6 (with 72 per cent of galaxies with $n$ in the range 1 to 2.5), representative of the full 51 objects. This suggests that, after we apply this correction to the F06 R$_{\rm e}$,s (R$_{\rm e,deV} = 0.77$R$_{\rm e,F06}$), the use of S\'ersic fit based R$_e$,s will not significantly bias M$_{\rm Gal,dyn}$ for the F06 objects. While we find good agreement for this small sample of galaxies for the specific methods used to determine R$_e$ by the respective authors, we caution that in general S\'ersic and R$^{1/4}$ based R$_{\rm e}$ typically show significant differences \citep{Trujillo:2001}.  After excluding disk-dominated galaxies of type Sb or later we were able to derive $M_\mathrm{Gal,dyn}$ for 48/86 galaxies; all objects for which a $\sigma$ and R$_e$ measurement was available. Typical errors on M$_{\rm Gal,dyn}$ are $\sim 50$ per cent.

\begin{figure*}[t]
\includegraphics[height=4.8in]{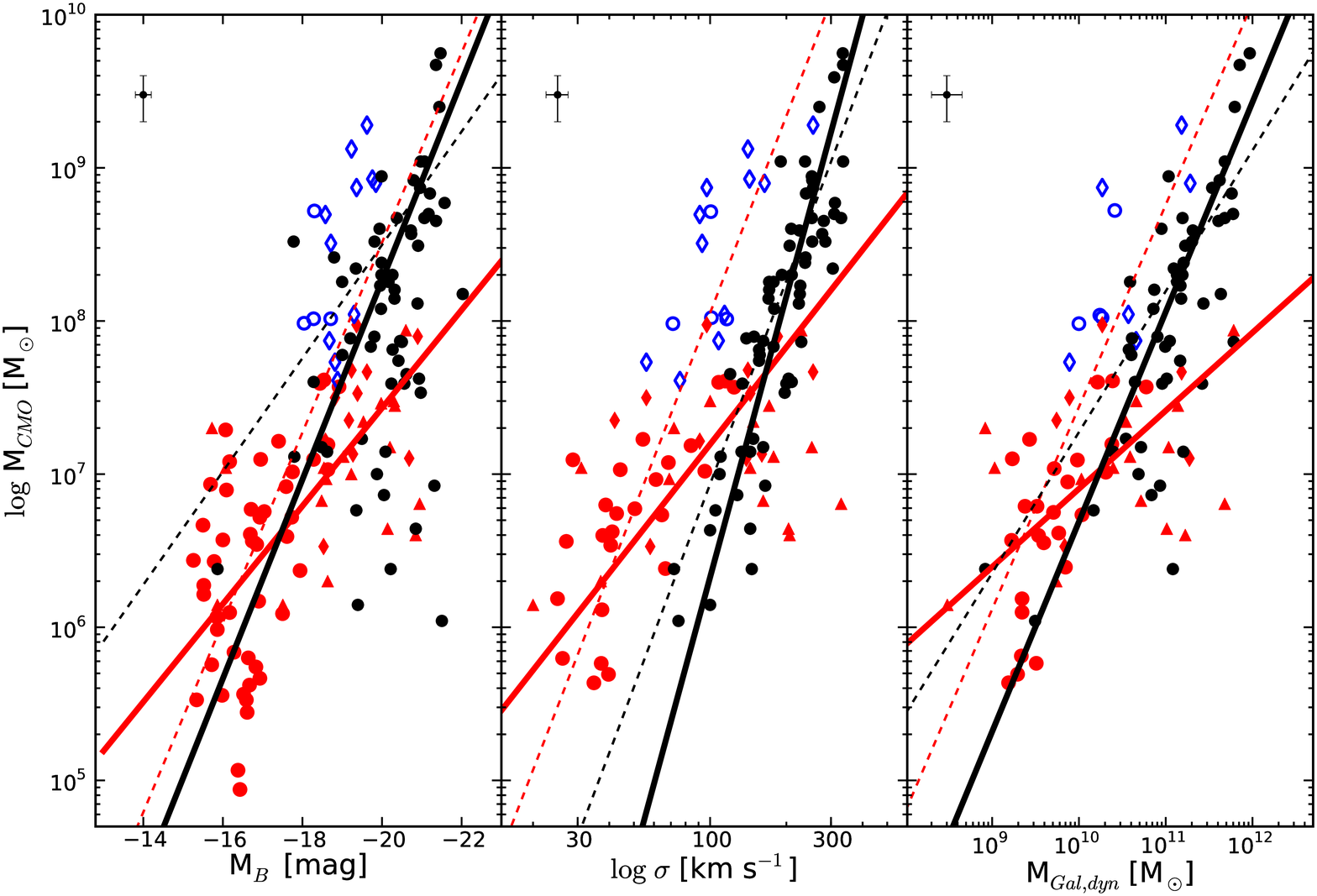}
\caption{M$_\mathrm{NC}$ and M$_\mathrm{BH}$ mass vs. galaxy magnitude $M_B$ (left panel), velocity dispersion $\sigma$ (middle panel) and dynamical mass M$_\mathrm{Gal,dyn}$ (right panel). Black dots indicate SMBHs, red symbols indicate NCs and open blue symbols show those objects identified as NDs. For the NCs and NDs the symbol indicates the sample each datapoint was drawn from: circles for F06, diamonds for BGP07 and triangles for GS09. The thick black and red lines indicate the best-fitting linear relations for the SMBH sample and the NC sample respectively. The thin dashed lines indicate the corresponding best-fitting relations from F06. A representative error bar is shown in the upper left corner of each panel. We note that lower-luminosity galaxies typically lie below the M$_{\rm BH}$--M$_B$ relation -- consistent with the bent M$_{\rm BH}$--M$_B$ relation shown by \citet{Graham:2012c} using the spheroid luminosity.} 
\label{fig:scaling1}
\end{figure*}

\begin{figure*}[t]
\centering
\includegraphics[height=4.8in]{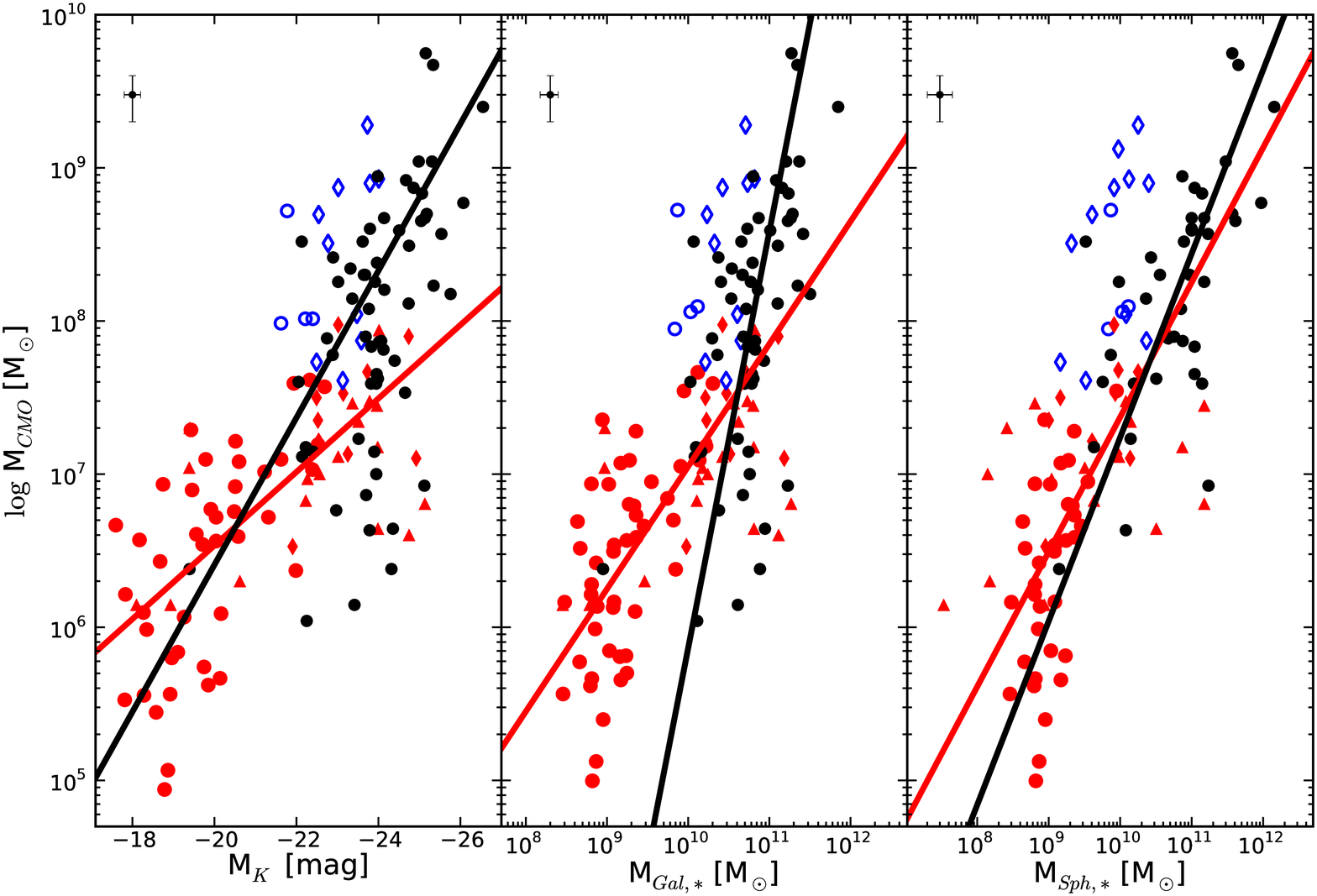}
\caption{M$_\mathrm{NC}$ and M$_\mathrm{BH}$ vs. galaxy {\it K-}band magnitude, M$_K$ (for galaxies classified E and dE only, left panel), galaxy stellar mass, M$_\mathrm{Gal,*}$ (middle panel) and spheroid stellar mass, M$_\mathrm{Sph,*}$ (right panel). Colors and symbols as in Figure \ref{fig:scaling1}. A representative error bar is shown in the upper left corner of each panel.}
\label{fig:scaling2}
\end{figure*}

We additionally determine stellar masses for the full galaxy, M$_\mathrm{Gal,*}$, and for the spheroidal component, M$_\mathrm{Sph,*}$. To determine M$_\mathrm{Gal,*}$ we multiplied the total galaxy luminosity of each object by the appropriate mass-to-light ratio. For all objects we used M$_K$ from 2MASS (excluding galaxies with no 2MASS M$_K$ total magnitude) and, following BGP07, assumed a standard mass-to-light ratio, M/L$_K = 0.8$ \citep[see also][]{Bell:2001}. M$_\mathrm{Gal,*}$ was determined for 80/86 galaxies. All magnitudes and colors were corrected for Galactic extinction following \citet{Schlegel:1998}. The data was not corrected for internal extinction, though we note that for most galaxies $M_K$ and $M/L_K$ are minimally affected by dust extinction. 

M$_\mathrm{Sph,*}$ was determined by multiplying the total {\it spheroid} magnitude of each object by the appropriate mass-to-light ratio as described above. We use the spheroid masses provided by GS09 for their galaxies. We adopt the same spheroid masses as BGP07, obtained by multiplying their {\it K}-band spheroid magnitudes by M/L$_K = 0.8$. For the F06 galaxies we use the galaxy stellar masses described above, excluding 16 galaxies classified as S0 or dS0 as they are likely to contain a large scale disk and no bulge-disk decomposition is available. This resulted in M$_\mathrm{Sph,*}$ for 66/86 galaxies -- all galaxies for which a spheroid mass or spheroid magnitude and an optical color were available. Typical errors on M$_{\rm Gal,*}$ are $\sim 25$ per cent and on M$_{\rm Sph,*} \sim 40$ per cent due to the increased uncertainty in separating the spheroid component of the galaxy's light.

\subsection{Supermassive black hole galaxy sample}
To compare to our nuclear star cluster sample, we take the supermassive black hole sample of \citet{Graham:2011}. This sample consists of 64 galaxies with directly measured supermassive black hole masses. The host galaxy velocity dispersions for this sample are presented in \citet{Graham:2012b}, the host galaxy {\it B-} and {\it K-}band luminosities in \citet{Graham:2012c} and the distance to each object in \citet{Graham:2011}.

We also determine derived quantities, M$_\mathrm{Gal,dyn}$, M$_\mathrm{Gal,*}$ and M$_\mathrm{Sph,*}$ for the supermassive black hole host galaxies following the approach for the nuclear star cluster host galaxies described above. Briefly, we derive M$_\mathrm{Gal,dyn}$ from the Virial estimator, using the velocity dispersions from \citet{Graham:2012b} and R$_e$ from the RC3. This allowed us to derive M$_\mathrm{dyn}$ for 40/64 galaxies. We derive galaxy stellar masses, M$_\mathrm{Gal,*}$ for the supermassive black hole galaxies as for the nuclear star cluster galaxies, using the galaxy {\it K-}band magnitude and an assumed M/L$=0.8$. We derive M$_\mathrm{Gal,*}$ for 59/64 galaxies -- all objects with an available {\it K-}band magnitude. To derive spheroid stellar masses we make use of the spheroid magnitudes, m$_\mathrm{Sph}$ presented in \citet{Marconi:2003}  \citep[with the exception of NGC 2778 and NGC 4564, see ][]{Graham:2007} and \citet{Haring:2004}\footnote{Recent works by \citet{Beifiori:2012}, \citet{Sani:2011} and \citet{Vika:2012} presented new bulge-to-disk decompositions for a significant number of galaxies in our SMBH sample. We elect not to make use of these new values for now because the agreement on the bulge-to-total flux ratios between the three authors is poor and it is unclear which provides the more accurate spheroid luminosities. For example, the three authors find bulge-to-total ratios of 0.78, 0.51 and 0.36 respectively for NGC 4596, with similar significant variations in spheroid luminosities.}. These were then multiplied by an appropriate M/L determined using the relations presented in \citet{Bell:2003}, with optical colors obtained from the HyperLeda database. We determined M$_\mathrm{Sph,*}$ for 39/64 SMBH galaxies -- all objects with an available m$_\mathrm{Sph}$ and optical color. 

\section{Analysis}
\label{sec:results}
\label{sec:scaling}
We use the {\sc BCES} linear fitting routine of \citet{Akritas:1996}, which minimizes the residuals from a linear fit taking into account measurement errors in both the $X$ and $Y$ directions. We adopt an orthogonal minimization, {\sc BCES(Orth)}, which minimizes the residuals orthogonal to the linear fit. An orthogonal regression provides a symmetrical treatment of the data, that is, swapping $x$ and $y$ data with each other still produces the same linear fit. This is preferred when one is after the underlying physical relation, referred to as the ``theorist's question" \citep{Novak:2006}. To compare the two sets of scaling relations it is important to use the same minimization technique because ``the different regression methods give different slopes even at the population level'' \citep{Akritas:1996}. We found that, in Monte Carlo simulations of a mock sample of NCs and SMBHs drawn from a single common CMO scaling relation, the {\sc BCES(Orth)} minimization proved most robust at recovering {\it the same} relation for both sets of datapoints. We therefore conclude that, to assess whether the two observed datasets are drawn from a single common scaling relation, the {\sc BCES(Orth)} minimization is the appropriate choice. We note that the {\sc BCES(Orth)} method is a common technique used in determining linear scaling relations and has been found to produce results consistent with other symmetric linear regressions \citep[e.g.][]{Haring:2004}. We additionally calculated the Spearman's rank correlation coefficient for each of the correlations reported in Table \ref{tab:fits}. With the exception of the M$_\mathrm{BH}$--M$_B$ relation, the probability that the given values of $r$ could arise if the quantities were not correlated is less than $0.01 \%$. For the M$_{\rm BH}$--M$_B$ relation this probability is $0.11 \%$.

\subsection{Nuclear star cluster mass scaling relations}
We have derived scaling relations connecting the nuclear star cluster mass to various properties of their host galaxies: {\it B-} and {\it K-}band luminosity, M$_B$ and M$_K$, velocity dispersion $\sigma$, galaxy dynamical mass M$_\mathrm{Gal,dyn}$ and galaxy and spheroid stellar masses, M$_\mathrm{Gal,*}$ and M$_\mathrm{Sph,*}$. These linear scaling relations are presented in Table \ref{tab:fits}. At present, given the relatively small samples and the significant errors on the derived nuclear star cluster masses, there is no compelling reason to fit more complicated broken or non-linear scaling relations to our data, though future studies with improved data may reveal additional complexity.Our Figure \ref{fig:scaling1} builds on Figure 2 from F06 by presenting linear fits of M$_\mathrm{NC}$ against: host galaxy {\it B}-band magnitude M$_B$, velocity dispersion $\sigma$, and virial mass M$_\mathrm{Gal,dyn}$. In Figure \ref{fig:scaling2} we present fits of M$_\mathrm{NC}$ against M$_K$, M$_\mathrm{Gal,*}$ and M$_\mathrm{Sph,*}$.

We find a slope of $2.11 \pm 0.31$ for the M$_\mathrm{NC} - \sigma$ relation. This is significantly shallower than that reported by F06 ($4.27 \pm 0.61$), though in better agreement with recently reported slopes of $1.57 \pm 0.24$ and $2.73 \pm 0.29$ \citep{Graham:2012a,Leigh:2012}. The principal difference between the F06 study and the  recent findings of a shallower slope of $\sim 2$ is the inclusion of nuclear star clusters in more massive galaxies with $\sigma > 200$ km$\ {\rm s}^{-1}$ (the F06 sample was limited to nuclear star clusters in host galaxies with $\sigma < 150$ km $\ {\rm s}^{-1}$). The exclusion of NDs from our fits (open blue symbols in all figures), which are typically an order of magnitude more massive than NCs, also contributes to our flatter slope, and accounts for the difference between our slope and that of \citet{Leigh:2012}.

We find a good correlation between M$_\mathrm{NC}$ and host galaxy luminosity in both the {\it B-} and {\it K-}band, with M$_\mathrm{NC} \propto L_K^{\sim 0.6 \pm 0.1}$. We find a strong correlation of M$_\mathrm{NC}$ with M$_\mathrm{Gal,*}$ and a somewhat weaker correlation with M$_\mathrm{Sph,*}$ -- this is consistent with the findings of \citet{Erwin:2012}, though we find a smaller difference between the strength of the correlations than they report(0.72 and 0.65, compared to their 0.76 and 0.38). 

We find a shallow slope of $\sim 0.5$ for the M$_\mathrm{NC}$ - M$_\mathrm{Gal,dyn}$ relation, which is significantly flatter than the slope $1.32 \pm 0.25$ reported by F06. We again attribute this difference to the inclusion of many more massive galaxies in our sample (though we caution that the Virial-estimator based dynamical masses used here and in F06 have significant errors). Bearing in mind that the relations were not constructed to minimise the scatter in the M$_{\rm CMO}$ direction, the M$_\mathrm{NC} - \mathrm{M}_\mathrm{Gal,dyn}$ relation has the lowest rms scatter (in the vertical M$_\mathrm{CMO}$ direction) of any of the NC scaling relations (though it is not significantly tighter than either the M$_\mathrm{NC} - \mathrm{M}_{K}$ or M$_{NC} - \mathrm{M}_\mathrm{Gal,*}$ relations).

\begin{figure*}[t]
\centering
\includegraphics[height=4.3in]{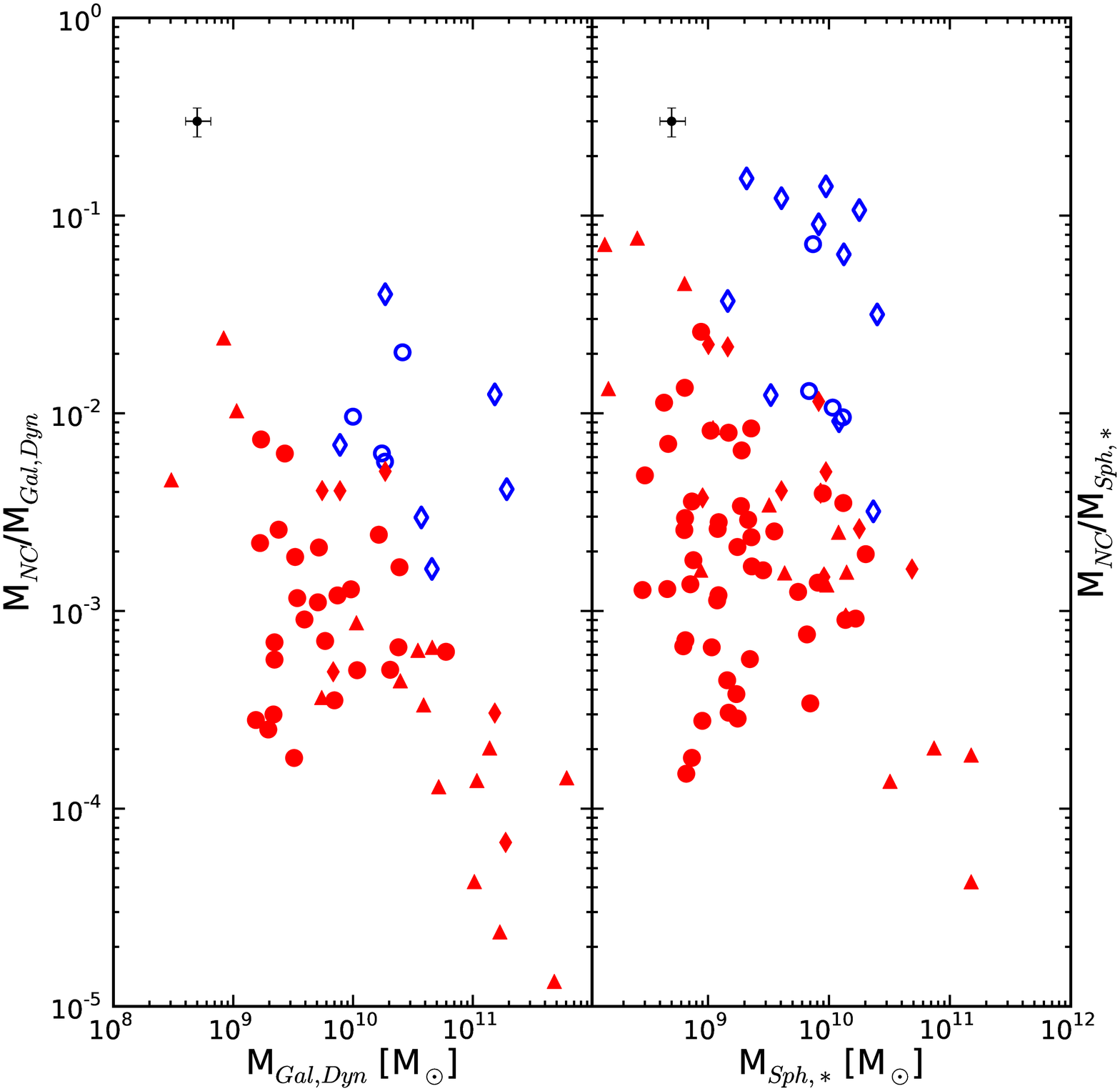}
\caption{Left panel: Nuclear star cluster mass, M$_\mathrm{NC}$ as a fraction of host galaxy dynamical mass, M$_\mathrm{Gal,dyn}$ vs. M$_\mathrm{Gal,dyn}$. Right panel: Same as left panel except using host spheroid stellar mass, M$_\mathrm{Sph,*}$. Colours and symbols as in Figure \ref{fig:scaling1}. The left panel shows a clear trend of decreasing CMO mass fraction with M$_\mathrm{Gal,dyn}$. While any trend is less clear with M$_\mathrm{Sph,*}$, we find that the CMO mass fraction still spans a large range of $\sim 2$ orders of magnitude. In both panels nuclear stellar disks (open symbols) are significantly offset to higher CMO mass ratios.}
\label{fig:mass_ratio}
\end{figure*}

\subsection{Supermassive back hole mass scaling relations}
In this subsection we derive a set of six scaling relations involving black hole masses. Except for the M$_\mathrm{BH} - \sigma$ relation, which involves galaxies of all types and is essentially a copy from \citet{Graham:2011}, due to available data these relations predominantly involve massive galaxies and spheroids. As such we have not included the developments which reveal a bent nature to the other five relations at lower masses \citep[e.g.][]{Graham:2012a,Graham:2012c}. However, these `bends' are such that the lower mass systems define steeper relations than shown here, which only emphasises the differences with the NC scaling relations discussed in the following section. 

We emphasise here that most of the relations we derive for supermassive black holes are {\it for comparison only} and do not represent the state-of-the-art in supermassive black hole scaling relations. For this reason we derive only simple linear fits to the available supermassive black hole sample for each variable. We do not distinguish between barred and unbarred galaxies \citep[see][for a discussion of the offset nature of barred galaxies in the SMBH-$\sigma$ relation]{Graham:2008a}, nor do we fit broken relations that better describe the scaling of supermassive black hole mass with host luminosity \citep{Graham:2012c} or mass \citep{Graham:2012a}. Whether theM$_\mathrm{BH}$--M$_\mathrm{Gal,*}$ and M$_\mathrm{BH}$--M$_\mathrm{Sph,*}$ relations are also bent is beyond the scope of this work but will be addressed in a future paper in this series.

The linear relations we derive are presented in Table. \ref{tab:fits} and are shown as the thick black lines in Figures \ref{fig:scaling1} and \ref{fig:scaling2}. While we emphasise again that these linear supermassive black hole scaling relations are for {\it comparison only}, we briefly discuss their consistency with similar scaling relations presented in the literature. We note that our M$_\mathrm{BH} - \sigma$ relation has a slope $\sim 6$, whereas a slope $\sim 4 - 5 $ had typically been reported in the literature \citep{Merritt:2001,Tremaine:2002}. However, our supermassive black hole sample now includes a significant number of barred galaxies which, as \citet{Graham:2008a} first observed, are offset from the unbarred relation. As \citet{Graham:2011} noted, including barred galaxies in ones sample increases the slope of the M$_\mathrm{BH} - \sigma$ relation -- \citet{Graham:2011} and \citet{Graham:2012b} report a slope of $5.95\pm 0.44$ and $5.76 \pm 1.54$ respectively for their full samples of both barred and unbarred galaxies,  consistent with our value. While the value we report is biased by the inclusion of the barred galaxies, their inclusion is appropriate as we do not distinguish between barred and unbarred galaxies in the corresponding M$_\mathrm{NC} - \sigma$ relation. 

The M$_\mathrm{BH}$ -- M$_\mathrm{Gal,dyn}$ and M$_\mathrm{BH}$ -- M$_\mathrm{Sph,*}$ relations we present, with slopes $\sim 1$, are typical of those reported in the literature \citep[e.g.][]{Marconi:2003,Haring:2004}. The correlation with spheroid stellar mass is significantly tighter than with total stellar mass, consistent with other studies \citep[e.g.][]{Kormendy:2001,McLure:2002}. Our relations are dominated by SMBHs with M$_\mathrm{BH} \gtrsim 10^8 M_{\odot}$, and \citet{Graham:2012b} has shown that above this rough threshold the M$_\mathrm{BH}$/M$_{\rm Gal,dyn}$ ratio is fairly constant, while at lower masses the M$_\mathrm{BH}$/M$_{\rm Gal,dyn}$ ratio is not constant but increasingly smaller, a result which can also be seen in our Figure \ref{fig:scaling1}, where the data points fall below the extrapolation of the solid black line at lower masses, causing the steepening of our M$_\mathrm{BH}$--M$_{\rm Gal,dyn}$ relation. We note that the correlation between M$_\mathrm{BH}$ and host galaxy luminosity  for the full sample is poor, with Spearmann r coefficients $\sim -0.4$. However, if we include only purely spheroidal systems (E and dE) this correlation improves markedly (Spearmann r = -0.81 and -0.70 in the {\it B-} and {\it K-}bands respectively), which is again unsurprising given that M$_\mathrm{BH}$ is known to correlate with the properties of the spheroid.

\section{Discussion}
\label{sec:discussion}
\subsection{Comparison of derived relations}
In contrast to F06 and WH06 we find that NCs and SMBHs do not follow common scaling relations. In all six diagrams that we have considered, the NCs and SMBHs appear to follow different relations (see Table \ref{tab:fits}). Our most significant finding is that the M$_{\rm NC}$--$\sigma$ relation (with a slope $\sim 2$) is significantly flatter than the M$_\mathrm{BH}$--$\sigma$ relation \citep[with a slope $\sim 6$ for all galaxies, or $\sim 5$ for barless galaxies:][]{Graham:2011}. This is in agreement with \citet{Graham:2012b}, but in contrast to F06 who find an M$_{\rm NC} - \sigma$ relation parallel to the M$_\mathrm{BH} - \sigma$ relation. The difference is due to: i) the exclusion of NDs from our NC sample; and ii) the inclusion of NCs that have masses higher than the SMBH/NC threshold of $10^7\ \mathrm{M}_\odot$ suggested by WH06. NDs are significantly more massive than NCs in comparable host galaxies and follow significantly different scaling relations \citep{Balcells:2007}. \citet{Scorza:1998} showed that NDs follow galaxy-scale stellar disk scaling relations, extending those relations to much lower mass.

In the middle panel of Figure \ref{fig:scaling1}, at a CMO mass of $\sim 10^{7}$ M$_\odot$, NCs are, on average, found in galaxies of significantly lower $\sigma$ and M$_\mathrm{Gal,dyn}$ than SMBHs of the same mass. 
\citet{Graham:2012a} has revealed that the M$_\mathrm{BH}$--M$_\mathrm{Gal,dyn}$ relation steepens from a slope of $\sim$1 at the high-mass end (M$_\mathrm{BH} \gtrsim 2\times10^8 M_{\odot}$) to a slope of $\sim$2 at lower masses (our slope of 1.37-1.55 is intermediate to these values because we fit a single linear relation to the high and low mass ends of what is a bent relation). The steepening in slope at the low-mass end is in the opposite sense to that observed for the NCs, which have a flatter slope of $0.55\pm0.15$. 

When considering the scaling of M$_\mathrm{CMO}$ with the stellar mass content of its host we find that M$_\mathrm{NC}$ appears to be driven by the total stellar mass, whereas M$_\mathrm{BH}$ is more closely associated with only the spheroidal component. Combining this finding with the result that M$_\mathrm{NC}$ follows much flatter relations with $\sigma$ and M$_\mathrm{Gal,dyn}$ than M$_\mathrm{BH}$ does, suggests that the physical processes that lead to the build-up of a nuclear stellar cluster may be significantly different to those that drive the formation of supermassive black holes. A complementary view is provided by Figure \ref{fig:mass_ratio}, where we plot the NC mass fraction as a function of host galaxy dynamical mass and spheroid stellar mass, i.e. M$_\mathrm{NC}$/M$_\mathrm{Gal,dyn}$ vs. M$_\mathrm{Gal,dyn}$ and  M$_\mathrm{NC}$/M$_\mathrm{Sph,*}$ vs. M$_\mathrm{Sph,*}$. The ratio M$_\mathrm{NC}$/M$_\mathrm{Gal,dyn}$ shows a clear trend with M$_\mathrm{Gal,dyn}$, in the sense that the NC mass fraction decreases smoothly with M$_\mathrm{Gal,dyn}$. We note that M$_\mathrm{NC}$/M$_\mathrm{Sph,*}$ is also {\it not} constant, spanning $\sim$2 orders of magnitude from 0.02 per cent to 2 per cent. 

\subsection{Galaxies hosting supermassive black holes and nuclear star clusters}
\label{sec:BH+NC_gals}

The GS09 sample of galaxies contain both an NC and an SMBH. It is likely that the SMBH and NC in these galaxies interacted in some way during their formation, hence additional physical processes may have influenced their scaling with their host galaxy. It is unclear what exactly the result of any interaction may be: it has been suggested that (i) the presence of a single SMBH may evaporate the NC \citep{Ebisuzaki:2001,OLeary:2006}, (ii) a binary SMBH may heat and erode the NC \citep{Bekki:2010}, and that (iii) some SMBHs are, in part, built up by the collision of NCs \citep{Kochanek:1987,Merritt:2004}. Additionally, given the existence of some dual-CMO galaxies, it is likely that some of the supposed NC- or SMBH-only galaxies in our sample contain an undetected SMBH or NC, respectively. Because of this it is unclear whether it is correct to include or exclude the GS09 galaxies from our main NC sample. More importantly, while including the GS09 galaxies does affect the NC scaling relations our conclusions do not depend on whether we include them.

\section{Conclusions}

We have revised three NC scaling relations and additionally presented three new scaling relations involving NC mass and host galaxy properties. We have also conducted a comparison of the scaling relations for NCs and SMBHs for the largest sample of objects to date. Our principal conclusions are:
\begin{enumerate}[i)]
\item The M$_\mathrm{NC}$--$\sigma$ relation is not parallel to the M$_\mathrm{BH}$--$\sigma$ relation when nuclear disks are properly identified and excluded and recent identifications of NCs in massive galaxies are included, in agreement with \citet{Graham:2012b}.
\item Nuclear star clusters and black holes do not follow a common scaling relation with respect to host galaxy mass, in agreement with BGP07.
\item The nuclear cluster scaling relations are considerably shallower than the corresponding supermassive black hole scaling relations. This is true for the relations involving host galaxy: $\sigma$, luminosity, dynamical mass and stellar mass. 
\item The dominant physical processes responsible for the development of NCs and SMBHs, in relation to their host galaxy or spheroid are suspected to be different given the above findings.
\item The NC mass fraction, with respect to the mass of its host galaxy or spheroid, is {\it not} constant, spanning $\sim$2 orders of magnitude. The NC mass fraction decreases in more massive galaxies. 
\end{enumerate}

\acknowledgements
We would like to thank Pat C\^{o}t\'{e} for his aid in deriving some of data used in this paper. This research was supported by Australian Research Council funding through grants DP110103509 and FT110100263. This research has made use of the NASA/IPAC Extragalactic Database (NED) which is operated by the Jet Propulsion Laboratory, California Institute of Technology, under contract with the National Aeronautics and Space Administration. We acknowledge the usage of the HyperLeda database (http://leda.univ-lyon1.fr)

\bibliographystyle{apj}
\bibliography{nuclear_clusters}

\end{document}